\documentclass[twocolumn,showpacs,amsmath,amstex,amssymb,mathfonts,prb]{revtex4-1}
\usepackage{graphicx,bm,units}
\usepackage{color}

\begin{document}

\title{Topological phonon modes in filamentary structures}

\author{Nina Berg, Kira Joel, Miriam Koolyk and Emil Prodan}
\address{Department of Physics, Yeshiva University, New York, NY 10016}

\begin{abstract}
This work describes a class of topological phonon modes, that is, mechanical vibrations localized at the edges of special structures that are robust against the deformations of the structures. A class of topological phonons was recently found in 2-dimensional structures similar to that of Microtubules. The present work introduces another class of topological phonons, this time occurring  in quasi one-dimensional filamentary structures with inversion symmetry. The phenomenon is exemplified using a structure inspired from that of actin Microfilaments, present in most live cells. The system discussed here is probably the simplest structure that supports topological phonon modes, a fact that allows detailed analysis in both time and frequency domains. We advance the hypothesis that the topological phonon modes are ubiquitous in the biological world and that living organisms make use of them during various processes. 
\end{abstract}


\date{\today}

\maketitle

Condensed matter research has been profoundly marked by the discovery of a new phase of matter called the topological insulating phase,\cite{kane2005A,kane2005B,bernevig2006c,koenig2007} which has already generated a plethora of amazing applications and changed the way we understand condensed matter.\cite{ZHassan2010,Qi2010} The hallmark of a topological insulator is the emergence of electronic states along any edge cut into a sample of such material. Although insulators when probed deep inside the bulk, these materials conduct electricity along their edges without any resistance. This property cannot be destroyed by any chemical or mechanical treatment of the edge.\cite{ProdanJMP2009,Prodan:2009mi} The phenomenon is not restricted only to electronic systems. Electromagnetic waves can display similar effects and it was shown that specially designed photonic crystals can support similarly robust electromagnetic wave-modes along their edges.\cite{Haldane:2008bf,Raghu:2008dm} Likewise, specially designed mechanical lattices can support topological phonon modes.\cite{Prodan2009wq}

Edge and surface phonon modes are not rare occurrences in hard matter systems. However, most of these modes are very sensitive to the deformations of the systems and they can be easily suppressed by various treatments of the edges or surfaces. There is a special effect that occurs in the presence of a magnetic field, the phonon Hall effect,\cite{StrohmPRL2005ff,InyushkinJETP2007vy} where robust vibrational edge modes can be observed very much like the edge electronic states in the Quantum Hall Effect. The effect was recently proposed to have a topological nature,\cite{ZhangPRL2010vu} by following an analysis similar to that of Ref.~\onlinecite{Prodan2009wq}. As opposed to the phonon Hall effect, the emergence of the topological phonon modes described in the present work (and in Ref.~\onlinecite{Prodan2009wq}) does not require any external field and is a consequence of the special intrinsic properties of the structures.  

The lattice presented in Ref.~\onlinecite{Prodan2009wq}, exhibiting topological phonon modes,  was inspired from the structure of Microtubules,\cite{Desai1997pd} a biomaterial synthesized by all living cells. Microtubules display a phenomenon known as dynamic instability (DI),\cite{Fygenson1994tj,Howard2003zs,Kueh2009yp, Dimitrov2008ty} in which they randomly grow and shrink in length, a process that is essential for their normal functioning. Most of today's chemo-therapies against cancer target Microtubules in an attempt to inhibit their DI.\cite{Jordan2004yt} For this reason, understanding the mechanism of DI is one of the most active research areas in biomedical research. Ref.~\onlinecite{Prodan2009wq} showed that the phonon spectrum of the dimer lattice of Microtubules displays Dirac cones like electrons do in graphene. It also showed that these Dirac cones can be split, leading to topological phonon modes, which were indeed observed in explicit calculations for ribbon geometries. It was then hypothesized that the topological edge modes play a major role in the DI.

Topological phonon modes may also be relevant to the dynamics of Microfilaments, which control cell motility by applying force on the cell membrane.\cite{Pollard2003ds, Kueh2009yp} They are made of the protein actin, arranged in a double-helical formation.\cite{Holmes2009xd} A pool of ATP-actin monomers is present in the cell, and actin polymerization draws on this pool. The ATP-monomers collide and bond with the ends of the existing filament branches, elongating them. The bound ATP-actin hydrolyzes into ADP-actin almost instantly, releasing quanta of about 12 kT energy.\cite{Gordon1976sd} At the same time, ADP-actin at the other end of the Microfilament network depolymerizes and returns to the pool as ATP-actin. Eventually capping proteins stop the growth of the branches.\cite{Pollard2003ds}

\begin{figure}
  \includegraphics[width=5cm]{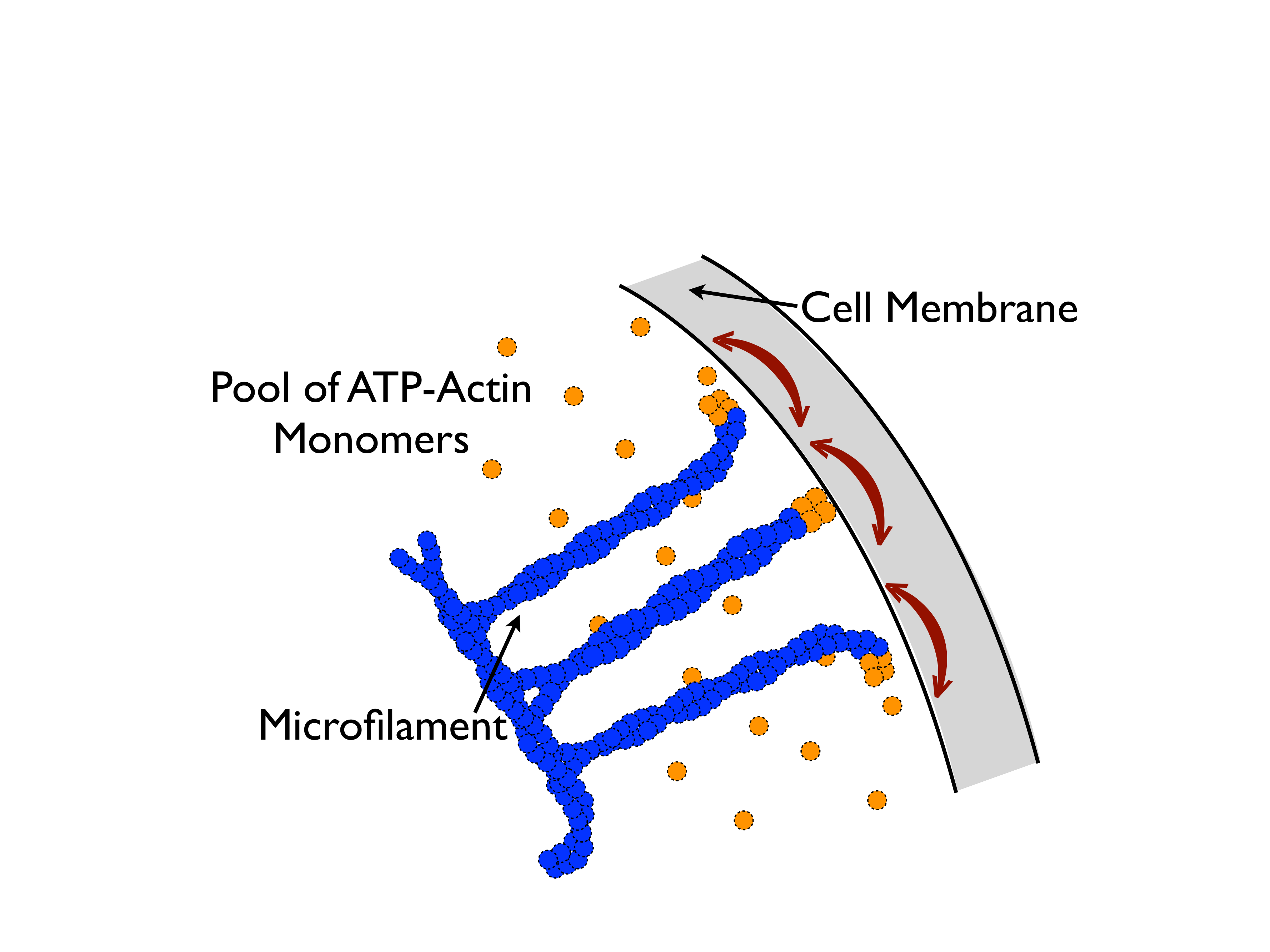}\\
  \caption{(Color online) The ends of the Microfilaments oscillate while pushing against the membrane, exposing their ends to the pool of available ATP-Actin. The ATP-Actin is then able to attach to the exposed ends and continue the filament's polymerization.}
 \label{Actin}
\end{figure}

One un-explained issue is the way in which the Microfilaments continue to grow while their ends push against the cell membrane, as is the case when they generate the motile force.\cite{Pollard2003ds} It is difficult to explain how the ATP-monomers from the solution are still able to collide with the ends of the filaments. According to the Elastic Brownian ratchet model,\cite{Mogilner1996gh, Pollard2003ds} the Microfilaments vibrate as spring-like wires and the edges adjacent to the cell membrane bend laterally, exposing the ends to the pool of ATP-actin. This allows additional actin monomers to squeeze in and attach themselves to the ends of branches, as illustrated in Fig.~\ref{Actin}. The restoring force straightens the Microfilament, which pushes against the cell wall generating the motile force. 

We hypothesize that the bending of the Microfilaments is caused by topological edge modes, powered by the 12 kTs released during hydrolysis of the ATP-actin. The present work demonstrates the existence of such modes in filamentary structures similar to that of the Microfilaments. As we shall see through explicit simulations, such edge modes do not allow the energy to dissipate into the bulk of the filaments, and could indeed lead to vigorous shakeup of the ends of the structures, even when excited with weak stimuli.  The phonon modes discussed in this work are of a different type from those previously found in Microtubules, which required a 2D structure and special interactions. We do not exclude that newly found modes may exist in Microtubules.

The paper gives a general classification of filamentary structures with inversion symmetry and provides a simple and direct criterium to identify those structures supporting topological phonon modes. The paper also presents an explicit mechanical lattice that displays this new phenomenon, which is investigated through analytic calculations and computer simulations in both time and frequency domains.

\section{A ${\bm Z}_2$ topological classification of the structures with inversion symmetry}

The topological classification of the electronic systems with inversion symmetry was recently discussed in Refs.~\onlinecite{Hughes2010gh} and \onlinecite{Turner2010cu}. Here we will adopt the 1 dimensional discussion presented in Ref.~\onlinecite{Hughes2010gh} to the context of mechanical waves. Consider a periodic 1D harmonic lattice, with $K$ degrees of freedom $\xi$ per repeating cell, governed by the following equations of motion:
\begin{equation}\label{EqMotion1}
\begin{array}{c}
\ddot{\xi}_n^\alpha=-\sum_{m,\beta} t^{m}_{\alpha\beta} \xi_{n+m}^\beta,
\end{array}
\end{equation}
where $t$'s are coefficients specific to each structure (see for example Eqs.~\ref{SmallOsc1} and \ref{SmallOsc2}).
The equations of motion for any periodic system slightly perturbed from its equilibrium configuration can be cast into the form shown in Eq.~\ref{EqMotion1}. The ansatz of traveling waves:
\begin{equation}
\xi_n^\alpha(t) = \mbox{Re}\big [ A^\alpha(k) e^{i(\omega t-k n)} \big ], \ k\in[-\pi,\pi]
\end{equation}
leads to the usual normal modes equation:
\begin{equation}\label{NormalModes}
[\hat{M}(k)-\omega^2]\vec{A}(k)=0,
\end{equation}
where $M(k)_{\alpha,\beta}=\sum_m t_{\alpha\beta}^m e^{-ikm}$ and $\vec{A}(k)$ is the $K$ dimensional vector encoding the amplitudes $A^{\alpha}(k)$. Let $\omega_s(k)$ and $\vec{{\cal A}}_s(k)$, $s$= $1,\ldots,K$, be the solutions to Eq.~\ref{NormalModes}. The solutions at $k$=$\pm\pi$ are identical. Also, $\vec{{\cal A}}_s(k)$ are assumed to be normalized to one: $\vec{{\cal A}}_s^*(k) \cdot \vec{{\cal A}}_s(k)$=1, for all $s$= $1,\ldots,K$.

One essential requirement is the existence of a gap in the vibrational spectrum of the system, that is, an interval $[\omega_-,\omega_+]$ empty of normal frequencies: $\omega_s(k)<\omega_-$ for $s=1,\ldots,S$ and $\omega_s(k)>\omega_+$ for $s=S+1,\ldots,K$, with $\omega_+$ strictly larger than $\omega_-$. We refer to the interval $[\omega_-,  \omega_+]$ as the spectral gap.

For each $k$, the $K$ normal modes $\vec{{\cal A}}_s(k)$ generate a $K$-dimensional complex space ${\bm C}^K$. The $S$ modes with frequencies below the spectral gap generate a $S$-dimensional subspace ${\cal S}$, which we should view as an $S$-dimensional hyperplane of the $K$-dimensional complex space. When the $k$-number is changed, this hyperplane twists, much as a Moebius band will in real 3-dimensional space. Our task is to classify the twists of the $S$-dimensional hyperplane.   

\subsection{General Arguments}

We assume that the 1D harmonic lattice has inversion symmetry, which implies the existence of a $k$-independent matrix ${\cal P}$ such that 
\begin{equation}\label{InvSym}
{\cal P}\hat{M}(k){\cal P}^{-1}=\hat{M}(-k),
\end{equation} 
where ${\cal P}$ is an unitary matrix that squares to indentity:
\begin{equation}
{\cal P} {\cal P}^\dagger = \hat{I}, \ \ {\cal P} {\cal P} = \hat{I}.
\end{equation}
The inversion symmetry and the matrix ${\cal P}$ will be discussed in great detail for our explicit example. In the following, we demonstrate that the harmonic lattices with inversion symmetry fall into at least two topological classes.

The following analysis is standard in Quantum Mechanics,\cite{Wilczek:1984bs} but is definitely not common for the present context. We need to define a parallel transport, that is, a rule that tells us how to change a vector in the hyperplane $S$ when $k$ is modified so that it stays parallel. The construction of the parallel transport is not unique, however, the emerging topological classification is independent of how the parallel transport is defined.\cite{Milnor1974xd} Let $\hat{P}(k)$ denote the projector onto the $S$-dimensional hyperplane generated by the normal modes with frequencies below the spectral gap:
\begin{equation}
\begin{array}{c}
[\hat{P}(k)]_{\alpha\beta}=\sum_{s=1}^S[\vec{{\cal A}}_s(k)]_\alpha [\vec{{\cal A}}_s(k)^*]_\beta.
\end{array}
\end{equation}
This is a $K$$\times$$K$ matrix and $S$-dimensional hyperplane ${\cal S}$ is then simply given by $\hat{P}(k){\bm C}^K$. A parallel transport can be defined by the so called monodromy $\hat{U}(k,k_0)$, defined as the unique solution to the equation:
\begin{equation}\label{AdTransp}
\begin{array}{c}
\partial_k \hat{U}(k,k_0) =i [\hat{P}(k),\partial_k \hat{P}(k)] \hat{U}(k,k_0),
\end{array}
\end{equation}
with the initial condition $\hat{U}(k_0,k_0)$=$\hat{P}(k_0)$. Once we computed this $\hat{U}(k,k_0)$ for each $k$$\in$$[-\pi,\pi]$, we can define the parallel transport as the map that takes an initial vector $\vec{{\cal A}}(k_0)$=$\sum_{s=1}^S c_s \vec{{\cal A}}_s(k_0)$ of the hyperplane at $k_0$ into the vector $\vec{{\cal A}}(k)$=$\hat{U}(k,k_0)\vec{{\cal A}}(k_0)$ of the hyperplane at $k$. Physically, this parallel transport gives the change in a vibrational state of the system when the frequency is adiabatically changed.

\begin{figure}
  \includegraphics[width=3cm]{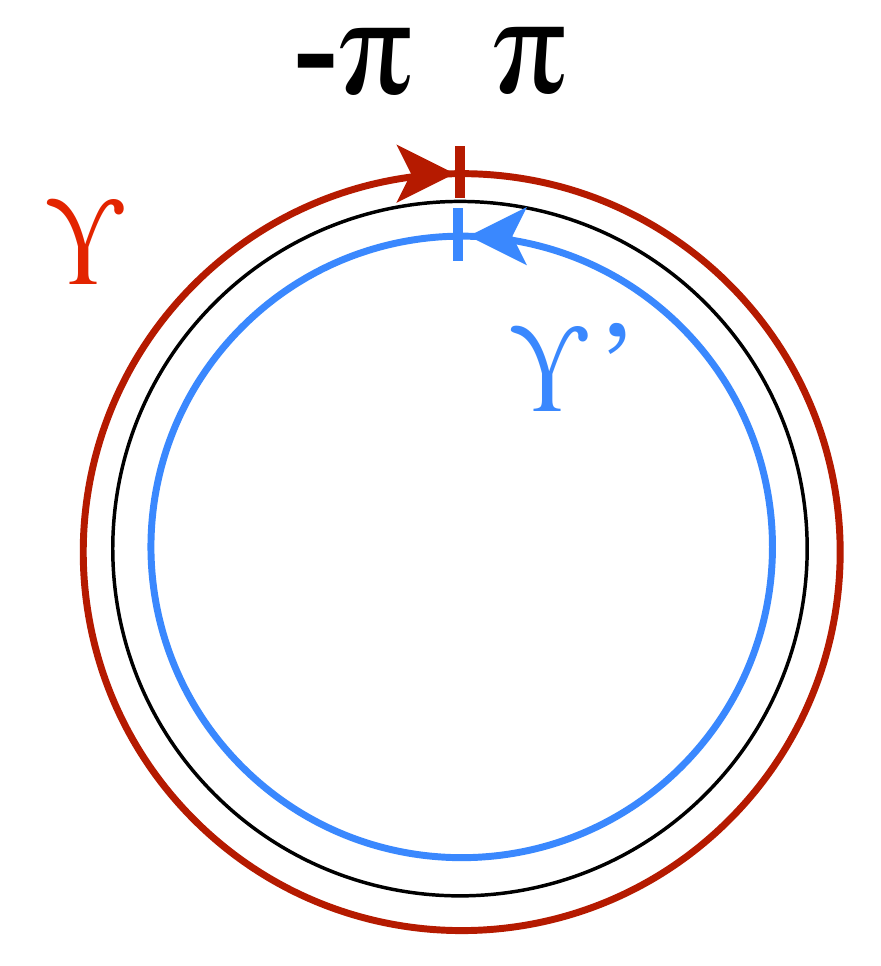}\\
  \caption{(Color online) The adiabatic transport is carried over the paths $\gamma$ and $\gamma'$.}
 \label{Paths}
\end{figure}

Since the normal modes at $k$=$\pm \pi$ are identical, we can identify these two $k$ points and think that $k$ is defined on a circle as  in Fig.~\ref{Paths}. If we do so, then we see that $\hat{U}(\pi,-\pi)$ takes the hyperplane $\hat{P}(-\pi){\bm C}^K$ into himself, becoming an unitary matrix that we will call $\hat{U}_{\gamma}$, to relate it to the paths shown in Fig.~\ref{Paths}. It is known from the non-Euclidean geometry that the parallel transport along a closed path will not necessarily return a vector into itself. Something similar happens here, that is, the map $\hat{U}_{\gamma}$ does not return a vector into itself.  

The monodromy $\hat{U}_{\gamma}$ has special properties when the inversion symmetry is present. Indeed, assuming $k_0$=$-\pi$, a conjugation of Eq.~\ref{AdTransp} with ${\cal P}$ gives:
\begin{equation}
\begin{array}{c}
\partial_k \{{\cal P}\hat{U}(k,-\pi){\cal P}^{-1}\} \medskip \\
=  i[\hat{P}(-k),\partial_k \hat{P}(-k)] {\cal P}\hat{U}(k,-\pi){\cal P}^{-1},
\end{array}
\end{equation}
with the initial condition ${\cal P}\hat{U}(-\pi,-\pi){\cal P}^{-1}$=$\hat{P}(\pi)$. This is just the equation for $\hat{U}(-k,\pi)$, which simply shows that ${\cal P}\hat{U}(k,-\pi){\cal P}^{-1}$ coincides with $\hat{U}(-k,\pi)$. Equivalently, we can think that ${\cal P}$ sends the path $\gamma$ into the path $\gamma'$ of Fig.~\ref{Paths}.  Now obviously $\hat{U}_{\gamma} \hat{U}_{\gamma'}$ equals identity, therefore:
\begin{equation}
\det \{\hat{U}_{\gamma} {\cal P}  \hat{U}_{\gamma} {\cal P}^{-1} \}=1.
\end{equation}
Using the elementary properties of the determinant, 
\begin{equation}\label{DetProp}
\det\{AB\}=\det\{A\}\det\{B\}, \ \det\{{\cal P}^{-1}\}=\det\{{\cal P}\}^{-1},
\end{equation}
we conclude that $\det \{ \hat{U}_{\gamma} \}^2$=1. Consequently, the determinant can take only two values:
\begin{equation}
\det \{ \hat{U}_{\gamma} \}=\pm 1.
\end{equation}

This is one of our main conclusions. It shows that the set of matrices $\hat{M}(k)$ satisfying Eq.~\ref{InvSym} can be split into two categories ${\cal C}_\pm$, the criterium being the value of $\det \{ \hat{U}_{\gamma} \}$ for the corresponding monodromy. A matrix $\hat{M}(k)$ that was placed in one category can not be morphed into a matrix from the second category by a continuous deformation that keeps the spectral gap open. Indeed, under such deformation, $\hat{U}_{\gamma}$ will change smoothly and, consequently, its determinant must change smoothly. Hence, it cannot make sudden jumps between $+1$ and $-1$. If the spectral gap closes, than $\hat{P}(k)$ becomes ill defined and the monodromy $\hat{U}_\gamma$ can no longer be defined. If the gap opens again, the monodromy $\hat{U}_\gamma$ can emerge with a different determinant.  

In the following, we describe how one can determine the signature of  $\det \{ \hat{U}_{\gamma} \}$ through an elementary calculation. We have successively:
\begin{equation}
\begin{array}{c}
\det \{\hat{U}_{\gamma} \} = \det \{ \hat{U}(\pi,0) \hat{U}(0,-\pi) \} \medskip \\
= \det \{ \hat{U}(\pi,0) {\cal P}\hat{U}(0,\pi) {\cal P}^{-1}  \}. \medskip \\
\end{array}
\end{equation}
Taking into account that ${\cal P}^{-1}$=${\cal P}$ and inserting the appropriate projectors, we obtain: 
\begin{equation}
\begin{array}{c}
\det \{\hat{U}_{\gamma} \} = \det \{ \hat{P}(\pi) \hat{U}(\pi,0) \hat{P}(0) {\cal P}\hat{P}(0)\hat{U}(0,\pi) \hat{P}(\pi) {\cal P}  \}.
\end{array}
\end{equation}
Using again the elementary properties of the determinant and the fact that $\hat{U}(\pi,0)\hat{U}(0,\pi)$=$Id$ we obtain:
\begin{equation}
\det\{ \hat{U}_{\gamma}\}=\det\{\hat{P}(0) {\cal P} \hat{P}(0)\} \det\{ \hat{P}(\pi) {\cal P}  \hat{P}(\pi) \}.
\end{equation}
In other words, all we need is to compute the action of inversion symmetry operation ${\cal P}$ at the special points $k$=0 and $k$=$\pi$. As we shall see, for concrete examples this can be accomplish through straightforward calculations. 

\subsection{Existence of the edge modes}

We show here that the systems with $\det\{ \hat{U}_{\gamma}\}$=$-1$ are likely to display edge phonon modes. For this, we imagine the following experiment. We take the infinitely long chain and we synchronously weaken the interactions between the 0th and 1st cell, and between the $N$th and $(N+1)$th cell, and so on, such that when these interactions are set to zero, we obtain decoupled finite chain pieces of length $N$. We will call the action we just described the decoupling process. During decoupling, we assume that the inversion symmetry is preserved.

\begin{figure}
  \includegraphics[width=8.6cm]{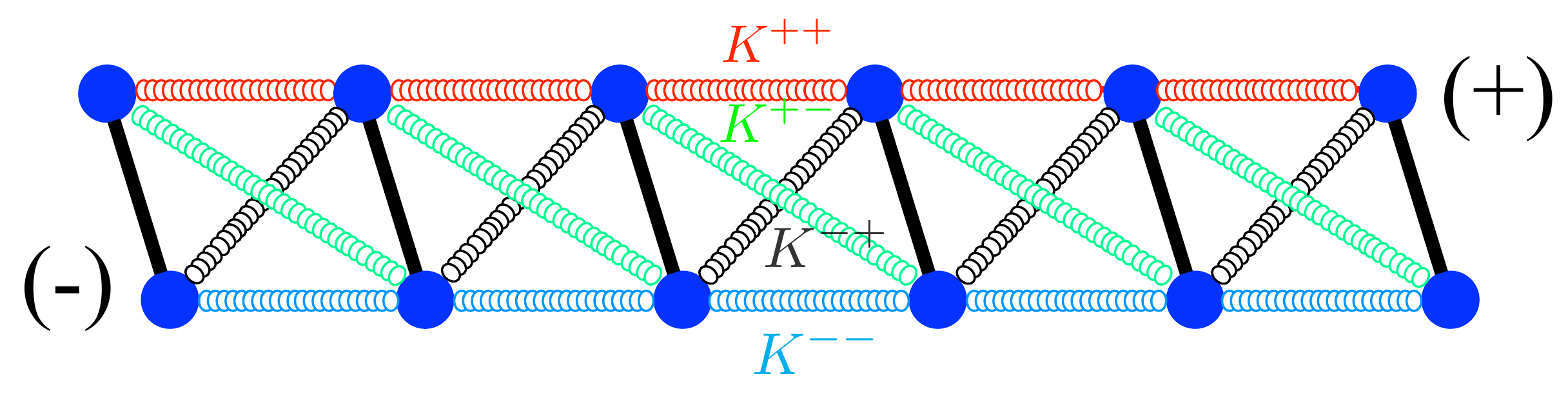}\\
  \caption{(Color online) The structure consists of a periodic array of dimers, connected by four different springs. The diagram depicts the system at equilibrium.}
 \label{System}
\end{figure}

At any moment of the decoupling process, we can see the altered chain as a new periodic chain whose repeating cell  contains $N$ cells of the original chain. Therefore the whole discussion of the last subsection  still applies. If one computes the monodromy matrix $\hat{U}_{\gamma}$ and its determinant for the new chain, he will find the following. At the beginning of the decoupling process, with the interactions untouched, $\det\{ \hat{U}_{\gamma}\}$=$\pm 1$ for chains belonging to ${\cal C}_\pm$ categories, respectively. At the end of the decoupling process, $\hat{M}(k)$ becomes independent of $k$ and consequently either $\hat{U}_{\gamma}$=$\hat{P}(-\pi)$ and $\det\{ \hat{U}_{\gamma}\}$=1, or $\hat{U}_\gamma$ cannot be defined because the system became gapless. For a system in the ${\cal C}_-$ category, both scenarios imply that the spectral gap closes during the decoupling. Thus, at least one pair of phonon bands must emerge in the spectral gap, one band coming from the upper and and one from the lower edge of the spectral gap. These two bands must move towards each other until they touch. In very special instances, which we were actually never able to observe, these two bands may continue to move until they disappear from the spectral gap. But in general, the bands end up inside the spectral gap at the end of the decoupling process, on top of each other. The bands become dispersionless, that is, completely flat, and the modes corresponding to these flat bands represent vibrational modes localized at the two ends of each finite piece of chain. This is in striking contrast with the case of simple harmonic lattices, that is, periodic arrays of harmonically interacting (point like) bodies without internal structure, where it was concluded in Ref.~\onlinecite{Jiang2010xu} that vibrational modes localized at the edges are very unlikely to occur in one dimensional structures.

The vibrational modes discussed above are robust, in the sense that, no matter how the decoupling is done, these edge modes will always emerge during the decoupling process. For example, a real lattice might become strongly distorted near the edges when a finite piece is being cut out but, even in this case, we still expect to see edge phonon modes for a system in the ${\cal C}_-$ category. 

\section{Example of a topological string}

In this section we introduce our example. It is the structure shown in Fig.~\ref{System}, made of an array of dimers, consisting of two masses $M$ linked by a rigid and massless arm of length $2d$. The adjacent dimers are connected by four different ideal springs. We will allow only two degrees of freedom per dimer: a horizontal motion of the center of mass, whose position will be marked by $x_n$ ($y_n$ will be constrained to 0), and a pivotal motion of the dimer in the plane of the structure and around the center of mass, described by the angle $\phi_n$. We assign the labels $\pm$ to the masses on the upper/lower rows of the structure. Using these labels, the springs between two dimers can be uniquely labeled by $\alpha\beta$, depending on which masses are connected by it. The corresponding spring constants will be denoted by $K_{\alpha\beta}$.

Let us point out that if we chose $K_{++}$=$K_{--}$, then the system is symmetric to the operations described in Fig.~\ref{InvSym}. These operations define the inversion symmetry of our system. 

\begin{figure}
  \includegraphics[width=8.6cm]{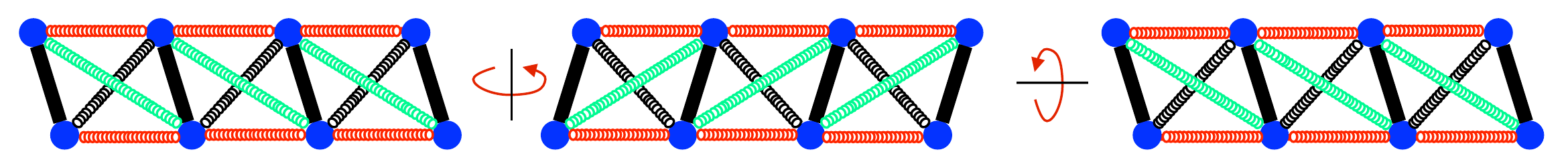}\\
  \caption{(Color online) If $K_{++}$=$K_{--}$, the structure is mapped into itself by the symmetry operations shown in this diagram.}
 \label{InvSym}
\end{figure}

\subsection{The Equations of Motion}

\begin{figure}
  \includegraphics[width=8cm]{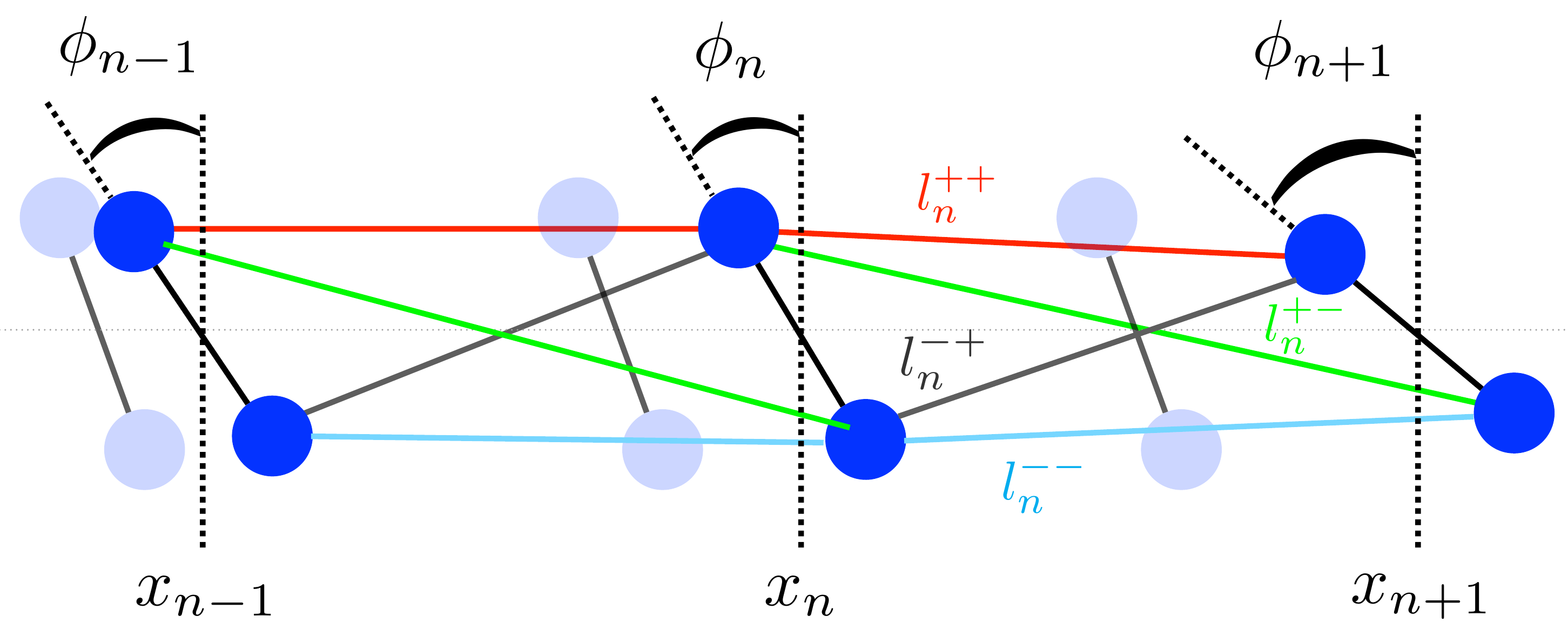}\\
  \caption{(Color online) The diagram introduces the notations used in the text. The array of dimers is shown at the equilibrium (faded) and at some arbitrary snapshot in time. $x_n$ represents the position of the center of mass of the $n$-th dimer and $\phi_n$ represents the angle of rotation of the $n$-th dimer relative to the vertical axis. In the equilibrium configuration, all $\phi_n$ are equal to $\phi_0$.}
 \label{Notations}
\end{figure}

We decided to discuss the equations of motion for the system in detail, in order to assist those who want to explore beyond what is presented here. With the notations introduced in Fig.~\ref{Notations}, the positions of two masses of the $n$-th dimer can be conveniently written as:
\begin{equation}
x_n^\alpha=x_n-\alpha d\sin(\phi_n); \ \ y_n^\alpha=\alpha d\cos(\phi_n).
\end{equation} 
The equations of motion for the $n$-th dimer are:
\begin{equation}\label{EqMo1}
2M \ddot{x}_n= F_{n,x}, \ \ 2Md^2 \ddot{\phi}_n = \tau_n,
\end{equation}
where $F_{n,x}$ and $\tau_n$ represent the net horizontal force and the net torque on the $n$-th dimer. They are given by:
\begin{equation}
\begin{array}{l}
F_{n,x}= \sum_{\alpha,\beta}\big[F_{n,x}^{\alpha\beta}-F_{n-1,x}^{\beta\alpha}\big] \end{array}
\end{equation}
and
\begin{equation}
\begin{array}{c}
\tau_n=\sum\limits_{\alpha,\beta}\big[(x_n^\alpha-x_n)(F_{n,y}^{\alpha\beta}-F_{n-1,y}^{\beta\alpha})\medskip \\
 - y_n^\alpha (F_{n,x}^{\alpha\beta}-F_{n-1,x}^{\beta\alpha})\big]
\end{array}
\end{equation}
where $F_{n,x/y}^{\alpha\beta}$ are the horizontal/vertical components of the forces generated by the springs at the right side of the $n$-th dimer:
\begin{equation}
\begin{array}{l}
F_{n,x}^{\alpha\beta}= K_{\alpha\beta} (1 - l_0^{\alpha\beta}/l_n^{\alpha\beta}) (x_{n+1}^\beta-x_n^\alpha)  \medskip \\
F_{n,y}^{\alpha\beta}= K_{\alpha\beta} (1-l_0^{\alpha\beta}/l_n^{\alpha\beta}) (y_{n+1}^\beta-y_n^\alpha),
\end{array}
\end{equation}
with:
\begin{equation}
l_n^{\alpha\beta}= \sqrt{(x_{n+1}^\beta-x_n^\alpha)^2 + (y_{n+1}^\beta-y_n^\alpha)^2}
\end{equation}
being the distance between the $\alpha$ and $\beta$ masses of the $n$-th and $n+1$-th dimers, and $l_0^{\alpha\beta}$ being the un-stretched length of the $\alpha\beta$ spring. 

The above equations of motion have been implemented numerically and the time evolution of the structure has been investigated using a 4th order Runge-Kutta time propagator. The results will be presented shortly. In all our calculations, we used length and time units so that $M=1$, $l_0^{++}=l_0^{--}=1$ and $d$ was set to 1.

\subsection{The Small Oscillations}

\begin{figure}
  \includegraphics[width=8.6cm]{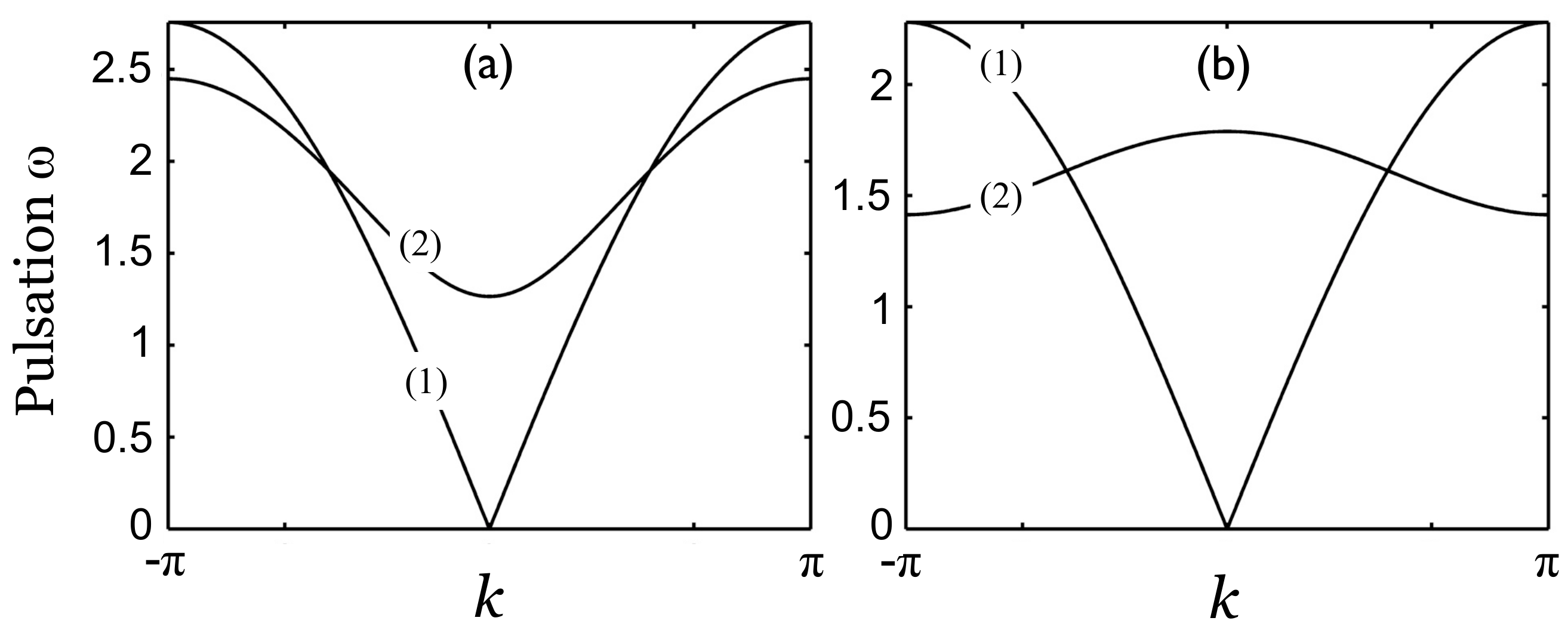}\\
  \caption{The phonon bands for (a) $K_{--}=K_{++}=0.5$ and $K_{-+}=K_{+-}=2$ and (b) $K_{--}=K_{++}=0.5$ and $K_{-+}=K_{+-}=4$.}
 \label{PhononSpec1}
\end{figure}

To explain and understand the features seen in our time domain simulations, it is useful to compute first the phonon spectrum of our structure, that is, the relation between the frequency and the wavenumber of the wave propagating modes. Let us introduce the notation: 
\begin{equation}
\xi^1_n=x_n-x_n^{\mbox{\tiny{eq}}}, \ \xi^2_n=d \phi_n.
\end{equation} 
In the linear regime, that is, when the dimers are only slightly perturbed from their equilibrium configurations, we have:
\begin{equation}
x_n^\alpha=x_n^{\mbox{\tiny{eq}}}+\xi^1_n-\alpha \cos(\phi_0)\xi^2_n, \ \ y_n^\alpha=-\alpha \cos(\phi_0)\xi^2_n.
\end{equation}
Plugging these expressions in Eq.~\ref{EqMo1} and retaining only the linear terms in $\xi^1$ and $\xi^2$, after tedious but otherwise straightforward calculations, we obtained the linearized form of  the equations of motion:
\begin{equation}\label{SmallOsc1}
\begin{array}{l}
-\ddot{\xi}_n^1=-\frac{1}{2}\sum\limits_{\alpha\beta} \Omega_{\alpha \beta}^{xx} (\xi_{n+1}^1-2\xi_n^1+\xi_{n-1}^1) \medskip \\
-\sum\limits_\alpha \alpha [ \Omega_{\alpha \alpha}^{xx} \cos \phi_0 + \Omega_{\alpha \alpha}^{xy} \sin \phi_0]\xi^2_n \medskip \\
+\frac{1}{2}\sum\limits_{\alpha \beta} \alpha [\Omega_{\alpha \beta}^{xx} \cos \phi_0 +\Omega_{\alpha \beta}^{xy} \sin \phi_0](\beta \xi_{n+1}^2 +\xi_{n-1}^2)
\end{array}
\end{equation}
and
\begin{equation}\label{SmallOsc2}
\begin{array}{l}
-\ddot{\xi}_n^2=-\sum\limits_\alpha \alpha [ \Omega_{\alpha \alpha}^{xx} \cos \phi_0 + \Omega_{\alpha \alpha}^{xy} \sin \phi_0]\xi^1_n \medskip \\
+\frac{1}{2}\sum\limits_\alpha \alpha [\Omega_{\alpha \beta}^{xx} \cos \phi_0 +\Omega_{\alpha \beta}^{xy} \sin \phi_0](\xi_{n+1}^1 + \beta x_{n-1}^1) \medskip \\
+\sum\limits_{\alpha\beta} \big \{ [\Omega_{\alpha \beta}^{xx} \cos^2 \phi_0 +\Omega_{\alpha \beta}^{yy} \sin^2 \phi_0] +\medskip \\  
\frac{\alpha}{2d}  
[ (F_{n,y}^{\alpha\beta}-F_{n-1,y}^{\beta \alpha}) \cos \phi_0 - (F_{n,x}^{\alpha \beta}  -F_{n-1,x}^{\beta \alpha}) \sin \phi_0] \big \} \xi_n^2 \medskip \\
-\sum\limits_{\alpha \beta} \beta [\Omega_{\alpha \beta}^{xx} \cos^2 \phi_0 +\Omega_{\alpha \beta}^{yy} \sin^2 \phi_0](\xi_{n+1}^2 + \xi_{n-1}^2).
\end{array}
\end{equation}
The coefficients have the following explicit expressions:
\begin{equation}
\begin{array}{c}
\Omega_{\alpha \beta}^{xx}=\frac{K_{\alpha \beta}}{M}\left \{1-\frac{l_0^{\alpha \beta}}{l_n^{\alpha\beta}} \left [ 1- \left ( \frac{\Delta x_{\alpha\beta}}{l_n^{\alpha\beta}} \right )^2 \right ] \right \}
\end{array}
\end{equation}
\begin{equation}
\begin{array}{c}
\Omega_{\alpha \beta}^{yy}=\frac{K_{\alpha \beta}}{M}\left \{1-\frac{l_0^{\alpha \beta}}{l_n^{\alpha\beta}} \left [ 1- \left ( \frac{\Delta y_{\alpha\beta}}{l_n^{\alpha\beta}} \right )^2 \right ] \right \} 
\end{array}
\end{equation}
and
\begin{equation}
\begin{array}{c}
\Omega_{\alpha \beta}^{xy}=\Omega_{\alpha \beta}^{yx}=\frac{K_{\alpha \beta}}{M}\frac{l_0^{\alpha \beta}}{l_n^{\alpha\beta}}  \frac{\Delta x_{\alpha\beta} \Delta y_{\alpha\beta}}{(l_n^{\alpha\beta})^2} ,
\end{array}
\end{equation}
with $\Delta x_{\alpha\beta}$=$x_{n+1}^\alpha$$-$$x_n^\beta$ and $\Delta y_{\alpha\beta}$=$y_{n+1}^\alpha$$-$$y_n^\beta$. The dimers are assumed in their equilibrium configuration in these last two equations.

To simplify the equations, we assume from now on that none of the springs are under tension at the equilibrium configuration and that the $l_0^{\alpha \beta}$ are chosen such that $\phi_0$=$0$. The ansatz $\xi_n^i $=$\mbox{Re}\{ A_i e^{i(\omega t-kn)} \}$ leads to the following normal modes equation $\hat{M}(k)\vec{A}=\omega^2 \vec{A}$, with:
\begin{equation}
\begin{array}{c}
\hat{M}(k)=\big [\sum_{\alpha\beta} K_{\alpha \beta}^{xx}-\sum_\alpha K_{\alpha \alpha}^{xx} \cos k \big ]\hat{I} \medskip \\
-\hat{\sigma}_1 (1 - \cos k)  \sum_\alpha \alpha K_{\alpha \alpha}^{xx} + \hat{\sigma}_2 \sin k  \sum_\alpha \alpha K_{\alpha -\alpha}^{xx} \medskip \\
- \hat{\sigma}_3 \cos k  \sum_\alpha \alpha K_{\alpha -\alpha}^{xx} ,
\end{array}
\end{equation}
where $\hat{\sigma}_i$'s are the Pauli's matrices:
\begin{equation}
\hat{\sigma}_1=\left ( ^0_1 \ ^1_0 \right ), \ \hat{\sigma}_2=\left ( ^0_i \ ^{-i}_{\ 0} \right ), \ \hat{\sigma}_1=\left ( ^1_0 \ ^{\ 0}_{-1} \right ).
\end{equation}
The normal modes equation has two solutions $\omega_{1,2}(k)$ and $\vec{A}_{1,2}(k)$.

Let us start the discussion of these solutions from the special case when $K_{++}$=$K_{--}$ and $K_{+-}$=$K_{-+}$. In this case, the normal modes equation becomes:
\begin{equation}
\begin{array}{c}
\left ( ^{\epsilon_1}_0 \ ^0_{\epsilon_2} \right )\left ( ^{A_1}_{A_2} \right )=\omega^2 \left ( ^{A_1}_{A_2} \right ),
\end{array}
\end{equation}
with
\begin{equation}
\begin{array}{c}
\epsilon_1=\sum_{\alpha\beta} \Omega_{\alpha \beta}^{xx}\big[1-\cos k \big ], \medskip \\
 \epsilon_2=\sum_{\alpha\beta} \Omega_{\alpha \beta}^{xx}\big[1-\alpha \beta \cos k \big ]. 
\end{array}
\end{equation}
The phonon modes decouple into a mode (corresponding to $\epsilon_1$) that involves displacements of the centers of mass but no pivotal motion and a mode (corresponding to $\epsilon_2$) which involves pivotal motion and no displacements of the centers of mass.

\begin{figure}
  \includegraphics[width=8.6cm]{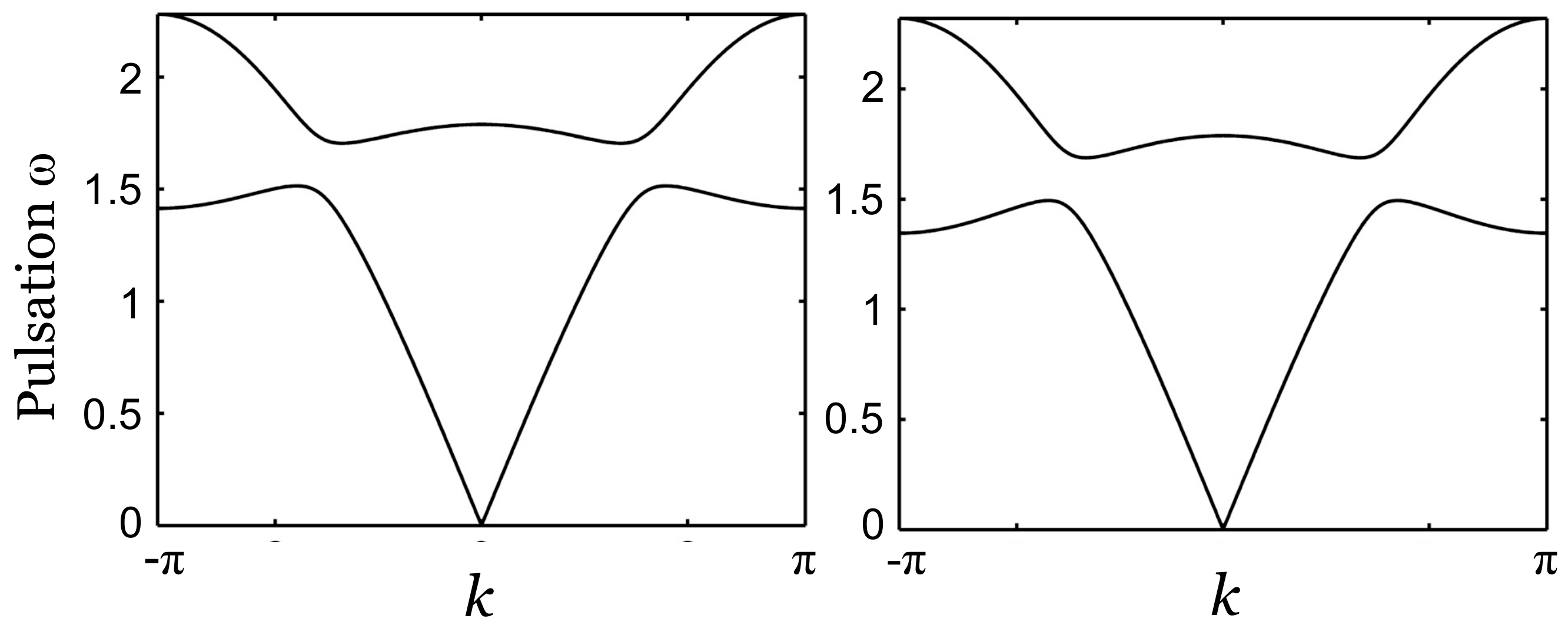}\\
  \caption{The spectral gap opening for (a) $K_{--}=K_{++}=0.5$, $K_{-+}=3$ and $K_{+-}=5$ and (b) $K_{--}=0.3$, $K_{++}=0.7$ and $K_{-+}=K_{+-}=4$.}
 \label{PhononSpec2}
\end{figure}

Depending on the common values given to $K_{++}$ and $K_{--}$ and to $K_{+-}$ and $K_{-+}$, the phonon bands can assume two distinct behaviors as shown in Fig.~\ref{PhononSpec1}. We are primarily interested in the case (b) because, as exemplified in Fig~\ref{PhononSpec2}, when the bands cross each other in that way, a small mismatch between $K_{+-}$ and $K_{-+}$ or between $K_{++}$ and $K_{--}$ will open a gap in the spectrum, thus realizing the last requirement of our general theory.

\subsection{The Edge Spectrum}

As already mentioned, if we set $K_{++}$=$K_{--}$, the system becomes symmetric to the inversion operation shown in Fig.~\ref{InvSym} and the $\hat{M}(k)$ matrix becomes:
\begin{equation}\label{MM}
\begin{array}{c}
\hat{M}(k)=\big [\sum_{\alpha\beta} K_{\alpha \beta}^{xx}-\sum_\alpha K_{\alpha \alpha}^{xx} \cos k \big ]\hat{I} \medskip \\
 + \hat{\sigma}_2 \sin k  \sum_\alpha \alpha K_{\alpha -\alpha}^{xx} 
- \hat{\sigma}_3 \cos k  \sum_\alpha \alpha K_{\alpha -\alpha}^{xx}.
\end{array}
\end{equation}
In the $k$-space, the inversion symmetry is implemented by ${\cal P}=\sigma_3$ and one can explicitly verify that:
\begin{equation}
\hat{\sigma}_3\hat{M}(k)\hat{\sigma}_3=\hat{M}(-k).
\end{equation}
Moreover, the projectors at the special points $k=0$ and $\pi$ can be computed explicitly. Assuming  $\sum_\alpha \alpha K_{\alpha -\alpha}^{xx}<0$, they are:
\begin{equation}
\begin{array}{c}
\hat{P}(0)=\frac{1}{2}(\hat{I}-\hat{\sigma}_3), \  \hat{P}(\pi)=\frac{1}{2}(\hat{I}+\hat{\sigma}_3)
\end{array}
\end{equation}
If $\sum_\alpha \alpha K_{\alpha -\alpha}^{xx}>0$, the expressions will be switched. In both cases, we can explicitly verify that:
\begin{equation}\label{Det}
\det\{\hat{P}(0) {\cal P} \hat{P}(0)\} \det\{ \hat{P}(\pi) {\cal P}  \hat{P}(\pi) \}=-1,
\end{equation}
and consequently the system is topologically nontrivial and we should observe edge modes. Note that when computing the determinants in Eq.~\ref{Det}, one needs to discard the null eigenvalues.

If we keep $K_{-+}=K_{+-}$ and consider a difference between $K_{--}$ and $K_{++}$, the system still has an inversion symmetry but in this case the symmetry is implemented in the $k$-space by the identity matrix. Hence:
\begin{equation}
\det\{\hat{P}(0) {\cal P} \hat{P}(0)\} \det\{ \hat{P}(\pi) {\cal P}  \hat{P}(\pi) \}=+1.
\end{equation}
Consequently, the system is trivial and we should not observe robust edge modes.

To verify these predictions, we performed the following numerical experiment. We considered a chain of 100 dimers linked in a circle configuration (the radius of the circle is very large so any physical bending of the structure can be ignored). In our theoretical argument of Section IB we used a gedanken experiment involving an infinite chain which was gradually sectioned in finite equal pieces. In practice, we cannot simulate such an infinite chain, but we can still start from a configuration with no edges (to avoid un-wanted phonon reflections), which is precisely the circle configuration mentioned above. Then we slowly weakened all four spring constants between two dimers, by making the substitution $K \rightarrow w K$ with $w$ continuously varying from 1 to 0. When $w=0$, the circle is fully opened in a finite piece with two separated edges, exactly like in the gedanken experiment. If topological phonon modes are present, we expect to see them gradually emerging from the bulk spectrum as described in Section 1B.

For each $w$ we have computed all 200 normal modes and we placed their frequencies on a vertical axes. Fig.~\ref{EdgeSpectrum} shows the results of these calculations, which illustrate how the frequencies of the normal modes change as $w$ is varied from 1 to 0. In the topological case $K_{--}=K{++}=0.5$ and $K_{+-}=3$ and $K_{-+}=5$ one can observe two solitary normal frequencies separating from the bulk spectrum and moving towards each other. When the two edges are completely formed at $w=0$, the two frequencies meet near the middle of the gap (the oscillatory motion of these modes is discussed in the next section). In contradistinction, no such modes are observed for the non-topological case $K_{--}=0.7$, $K_{++}=0.3$ and $K_{-+}=K_{+-}=0.5$. This explicit calculation confirms our general theoretical predictions.

\subsection{Time Domain Analysis}

\begin{figure}
  \includegraphics[width=8.6cm]{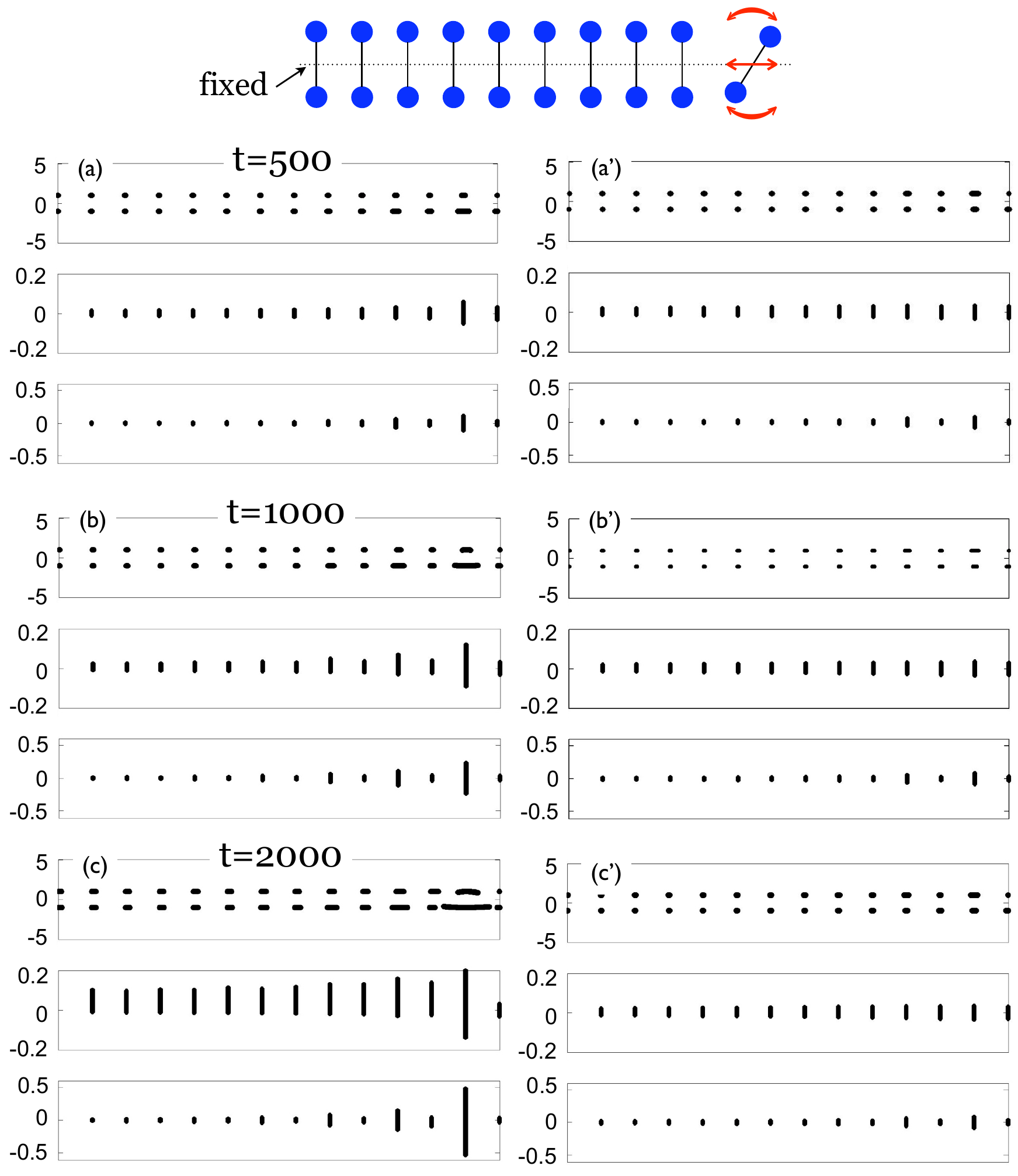}\\
  \caption{(Color online) Top diagram: Representation of the dimer chain and the forced motion of the rightmost dimer. Main panel:  The trajectories covered by the dimers after $t$=500, 1000 and 2000. Each sub-panel contains three plots, representing, from the top to bottom, the actual trajectories of the dimers, the trajectory of the horizontal displacements $x_n(t)$, and of the angular displacements $\phi_n(t)$. Plots (a) (b) and (c) represent the behavior of the dimers in the topological case where  $K_{--}=K{++}=0.5$ and $K_{+-}=3$ and $K_{-+}=5$. Plots (a$^\prime$) (b$^\prime$) and (c$^\prime$) represent the behavior of the dimers in the non-topological case where $K_{--}=0.7$, $K_{++}=0.3$ and $K_{-+}=K_{+-}=4$.}
 \label{EdgeResonance}
\end{figure}

As already mentioned, the (non-linear) equations of motion have been implemented numerically and the motion of the dimmers has been studied in real time \footnote{See EPAPS Document No. XXX for a demonstration}. Here, we discuss the manifestation of the edge modes in the real time dynamics of the dimer chain. For this, we considered a chain containing $N$=100 dimers plus two additional "edge" dimers. We fixed the very left dimer to the upward position while forcing the very right dimer into the following motion:
\begin{equation}\label{ForcedMotion}
x_{N+1}  = A_1 \sin \omega t, \ \ \phi_{N+1}=A_2 \sin \omega t.
\end{equation}
The rest of the $N$ dimers were released from the equilibrium with zero velocities (see Fig.~\ref{EdgeResonance}).

\begin{figure}
  \includegraphics[width=8.6cm]{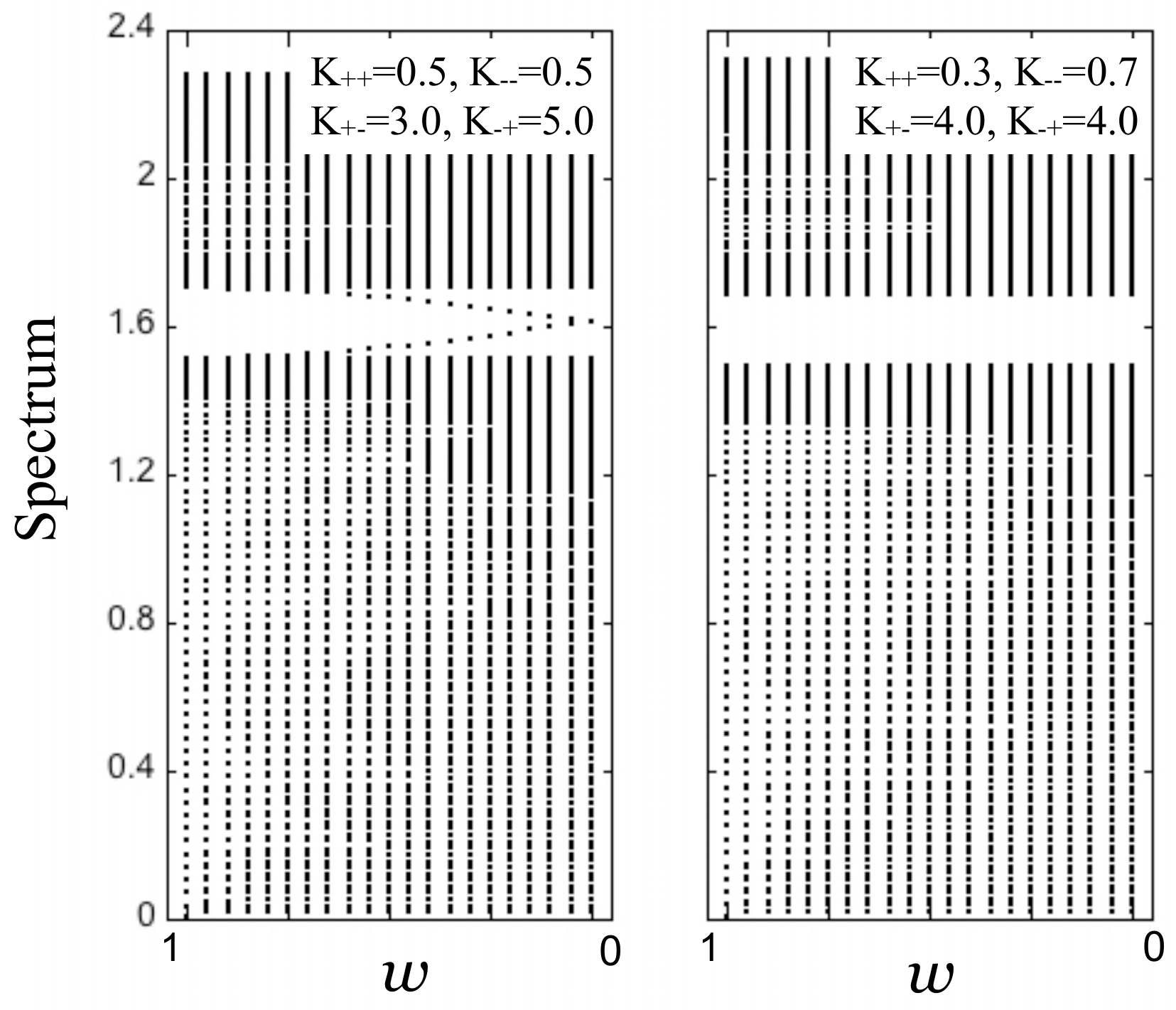}\\
  \caption{Each vertical sequence of dots represent the frequencies of the normal modes of a chain made of 100 dimers arranged in a circle. The spring constants $K$ of the springs connecting two dimers were gradually weakened by making the substitution $K$$\rightarrow$$wK$ and letting $w$ vary from 1 to 0. The frequencies were recomputed for many $w$ values. Panel (a) refers to the topological case, and panel (b) to the non-topological case. }
 \label{EdgeSpectrum}
\end{figure}

The resulting motion of the dimers has been plotted every $\Delta t$=3, and movies were generated for various pulsations $\omega$. For a more complete representation, we chose to look at (1) the actual dimer positions, (2) only at the horizontal displacement $x_n(t)$ of each center of mass, and (3) only the angular displacement $\phi_n(t)$ of the dimers as a functions of time. In the following, we will make reference to the wave modes discussed in subsection IIC. The movies (see the supplemented material) reveal that, indeed, if $\omega$ is below approximately 1.4 (see Fig.~\ref{PhononSpec1}), only the wave propagating mode that involves the translation of the dimers is excited, even though $A_2$ was given the same value of 0.03 as $A_1$. Above 1.4 and below the edge of the spectral gap we see both wave modes being excited. Similar behaviors are observed above the spectral gap. What really interests us is what happens when $\omega$ takes values inside the spectral gap. In order to generate a meaningful plot, we let the movies progress without erasing the images already played out. The resulting plots shows the trajectories followed by each dimer and, in particular, they reveal the amplitudes of the oscillating motion for each dimer at the time when the movie was stopped. In Fig.~\ref{EdgeResonance}, the last dimer was forced into the motion described in Eq.~\ref{ForcedMotion} with $A_1$=$A_2$=0.3 and $\omega$=1.6, chosen to be in the middle of the spectral gap. The movie was allowed to run for three time intervals: 500, 1000, 2000. 

The left panels in Fig.~\ref{EdgeResonance} simulate the situation illustrated in Fig.~\ref{EdgeSpectrum}(a). This is the topological case, for which the spring constants were set to $K_{--}=K_{++}=0.5$ and $K_{+-}=3$ and $K_{-+}=5$ and an edge resonance is expected. As one can see, only the amplitudes of the dimers close to the right edge reach appreciable values and the amplitudes are seen to decay exponentially away from the edge, demonstrating that we are indeed dealing with an edge mode. Furthermore, one can see the amplitudes increasing as time progressed, the angular displacement of the second right dimer reaching an amplitude of approximately 0.4 after $t=2000$, more than 10 times the amplitude of the forced oscillation imposed on the first dimer. This reveals an important property of the edge resonance, namely the ability to absorb and store energy in the proximity of the edge of the structure. 

The right panels in Fig.~\ref{EdgeResonance} simulate the non-topological case presented in Fig.~\ref{EdgeSpectrum}(b), where the spring constants were set to $K_{--}=0.7$, $K_{++}=0.3$ and $K_{-+}=K_{+-}=4.0$. In this case we don't expect an edge mode and indeed the amplitudes of the dimers near the edge remain practically zero at all times.

There is a small artifact in the plots of Fig.~\ref{EdgeResonance}. Due to a non-stationary effect steaming from the relatively large amplitude used for shaking the end dimer, the average positions of the centers of the dimers slowly drift to the left in time. Because of the way we generate Fig.~\ref{EdgeResonance}, this drift may give the impression of a finite oscillation amplitude for dimers away from the edges. We prompt the reader to watch the movie, where one can clearly see that the oscillation amplitudes of the dimers away from the edges are practically zero (the slow drifting motion is also visible).  

\section{Conclusions}
Using a monodromy argument, we have demonstrated that the filamentary structures with inversion symmetry fall into at least two topologically distinct classes. We have argued that one of the two classes contain systems that should display robust topological edge modes. The general theoretical predictions were verified using an explicit example of a mechanical structure that display robust edge modes. 

The structure was inspired from that of actin Microfilaments. We hypothesized that the energy from the ATP hydrolysis in Microfilaments is stored in these robust vibrations and is then used to drive the motion described in the Elastic Brownian ratchet model. This opens an interesting research direction, which is worthwhile pursuing because we believe that, with the present technology,  well designed and focused experiments can reveal if the Microfilaments have or have not edge modes.

Although not very accurate, the model allowed us to present the concept of topological phonon modes in a very explicit and detailed exposition, without being derailed by unnecessary complications. The simplicity of the structure will allow detailed analyses of other interesting questions, such as what happens when defects are present throughout the bulk of the chain, how is the edge excited by the hydrolysis of GTP actin or tubulin, how is damping effecting the picture, etc.. The structure presented here can become a very useful pedagogical tool to introduce the concept of topological phonon modes in an accessible and explicit way via computer simulations or real lab observations.     

\begin{acknowledgments} 
This research was supported by a Cottrell award from the Research Corporation for Science Advancement and by the office of the Provost of Yeshiva University. 
\end{acknowledgments}


%

\end{document}